\documentclass[aps,preprint,superscriptaddress]{revtex4-2}
\usepackage{amsmath, amsfonts, amssymb, amsthm, mathtools}  % For math
\usepackage{physics, empheq}  % For advanced expressions
\usepackage{hyperref}  % For links
\hypersetup{
  colorlinks=true,
  linkcolor=magenta,
  citecolor=blue,
  urlcolor=blue
}
\usepackage{graphicx}
\graphicspath{
  {fig/},
  {/home/raraki/github/TikZ_figures/},
  {/home/raraki/Simulation/analysis/result/3DTG/},
  {/home/raraki/Simulation/3DTG/}
}
\usepackage[svgnames]{xcolor}  % For additional colors
\usepackage{siunitx}  % For values with units
\usepackage[most]{tcolorbox}  % For annotations
\begin{document}

%\preprint{}

\title{
  Space-local Navier--Stokes turbulence
}


\author{Ryo Araki\footnote{Current address:
  Department of Mechanical and Aerospace Engineering,
  Faculty of Science and Technology,
  Tokyo University of Science,
  Yamazaki 2641, Noda-shi
  278-8510, Japan}}
\email[]{araki.ryo@rs.tus.ac.jp}
%\thanks{}
\affiliation{
  Univ Lyon, \'Ecole Centrale de Lyon, CNRS, Univ Claude Bernard Lyon 1, INSA Lyon,
  LMFA, UMR5509, 69130, \'Ecully, France
}
\affiliation{
  Graduate School of Engineering Science,
  Osaka University,
  1-3 Machikaneyama, Toyonaka, Osaka 560-8531, Japan
}
\author{Wouter J.~T.~Bos}
\affiliation{
  Univ Lyon, \'Ecole Centrale de Lyon, CNRS, Univ Claude Bernard Lyon 1, INSA Lyon,
  LMFA, UMR5509, 69130, \'Ecully, France
}
\author{Susumu Goto}
\affiliation{
  Graduate School of Engineering Science,
  Osaka University,
  1-3 Machikaneyama, Toyonaka, Osaka 560-8531, Japan
}


\date{\today}

\begin{abstract}
%%%%% Introduction %%%%%
We investigate the physical-space locality of interactions in three-dimensional incompressible turbulent flow.
%%%%% Details of motivation and previous studies %%%%%
%%%%% Research question %%%%%
%%%%% Approach %%%%%
To that, we modify the nonlinear terms of the vorticity equation such that the vorticity field is advected and stretched by the locally induced velocity.
This \emph{space-local} velocity field is defined by the truncated Biot--Savart law, where only the neighboring vorticity field in a sphere of radius \(R\) is integrated.
%%%%% Results 1 & 2 %%%%%
We conduct direct numerical simulations of the space-local system to investigate its statistics in the inertial range.
We observe a standard \(E(k) \propto k^{-5/3}\) scaling of the energy spectrum associated with an energy cascade for scales smaller than the space-local domain size \(k \gg R^{-1}\).
This result is consistent with the assumption \citet{Kolmogorov1941_the_local} made for the space-locality of the nonlinear interactions.
The enstrophy amplification is suppressed for larger scales \(k \ll R^{-1}\), and for these scales, the system exhibits a scaling consistent with a conservative enstrophy cascade.
%%%%% Significance and impact %%%%%
\end{abstract}

%\keywords{}

\maketitle

% % Table of contents
% \clearpage
% \tableofcontents

% % Fix ToC-generating commands
% % cf. https://tex.stackexchange.com/a/37392
% \makeatletter
% \let\toc@pre\relax
% \let\toc@post\relax
% \makeatother
% %
% % List of figures/tables
% \listoffigures
% \listoftables

\clearpage
\section{
  Introduction
  \label{sec:Introduction}
}
% \subsection{}
% \subsubsection{}

% ++++++++++++++++++++++++++++++++++++++++++++++++++
% \subsection*{Scale-space locality of turbulence}
% ++++++++++++++++++++++++++++++++++++++++++++++++++

Kolmogorov postulated that in turbulent flows at large Reynolds numbers, universal statistics should emerge at scales sufficiently smaller than the scale determined by boundary condition and larger than the dissipative scale~\citep{Kolmogorov1941_the_local}.
The origin of this universality is the scale locality of the nonlinear interactions.
It is a common concept in the turbulence community, as \citet[p.~104]{Frisch1995_turbulence} states in a footnote that: ``In turbulence, `local' and `localness' usually refer to scales, not to positions as in other areas of physics.''

% ++++++++++++++++++++++++++++++++++++++++++++++++++
% \clearpage
% \subsection*{Research question}
% ++++++++++++++++++++++++++++++++++++++++++++++++++

The concept of scale locality states that the collective effect of nonlinear interactions involving disparate length scales is weak compared to that of comparable length scales.
In fact, Kolmogorov developed his theory in physical space; he assumed that when turbulence is locally homogeneous, isotropic, and stationary in sufficiently a small domain far away from the boundaries, the system should exhibit universal statistics.
If we restrict the analysis to a small subdomain, interactions are implicitly assumed to be local in space.
Indeed, vortices separated beyond an appropriately defined correlation length are not expected to interact strongly.
However, this property has received little attention compared to the concept of scale locality.
For a concise review, see~\citet[\S~1.3.5 and \S~6.2.2]{Tsinober2009_an_informal}.
How do spatially local and nonlocal nonlinear interactions contribute to the small-scale universality of turbulence?
In other words, how is space locality related to scale locality?
How important are interactions between nearby vortices of different sizes compared to those between vortices of similar scales but far away?
The present study considers these research questions in three-dimensional incompressible Navier--Stokes turbulence.

% ++++++++++++++++++++++++++++++++++++++++++++++++++
% \clearpage
% \subsection*{Physical-space locality of turbulence}
% ++++++++++++++++++++++++++++++++++++++++++++++++++

The physical-space locality of the energy flux was first considered by~\citet{Kraichnan1974_on_Kolmogorovs}.
Since then, several formulations of the space-local energy flux have been proposed; Meneveau employed wavelets~\citep{Meneveau1991_analysis} and Lagrangian correlation framework to locally track the energy flux in space~\citep{Meneveau1994_on_the_Lagrangian}.
Eyink used a refined similarity hypothesis involving a spatial length scale~\citep{Eyink1995_local}, as well as developing a multi-scale gradient expansion, which decomposes the turbulent stress tensor into multi-scale and multi-order spatial derivatives~\citep{Eyink2005_locality, Eyink2006_multi-scale, Eyink2009_localness}.  % , Aluie2009_localness
Tsinober discussed spatially concentrated vorticity and its nonlocal interactions with background turbulence~\citep{Tsinober1998_is_concentrated}.
``Five-dimensional'' (three-dimensional space, scale, and time) analysis of energy cascade revealed an emergence and disappearance of fluid structures within the larger- and smaller-scale structures, respectively~\citep{Cardesa2017_the_turbulent}.
\citet{Doan2018_scale} investigated the scale-local energy cascade in terms of vortex stretching in real space.
More recent work found a power-law correlation between the filtered strain rate and the space-local energy flux~\citep{Alexakis2020_local}.
In~\citet{Vela-Martin2021_entropy}, the irreversibility of turbulence and preference for direct energy cascade is discussed with its space-local property.
We note that these questions are not only of fundamental importance but are also relevant to turbulence modeling, particularly in refining sub-grid scale models~\citep{Borue1998_local}.

Pressure and its Hessian are also important to describe the physical-space nonlocality of turbulence~\citep{She1991_structure}.
Several investigations report its role in decaying turbulence~\citep{Kishiba1993_physical-space}, relation with finite time blow-up~\citep{Ohkitani1995_nonlocal, Chae2022_on_a_type}, rotation~\citep{Nomura1998_the_structure}, and the role of local and nonlocal contributions to the influence of the pressure Hessian~\citep{Chevillard2011_local}.

Hamlington \textit{et al.} proposed a space local/nonlocal decomposition of the velocity field to investigate the alignment between the vorticity and the eigenvectors of the strain-rate tensor~\citep{Hamlington2008_local, Hamlington2008_direct, Hamlington2009_physics-based}.
% They revealed that the most extensional eigenvector of the nonlocal strain-rate tensor aligns with vorticity.
Later, Buaria \textit{et al.} used the same decomposition and observed self-attenuation of intense vorticity due to the locally induced strain rate~\citep{Buaria2020_self-attenuation, Buaria2021_nonlocal}.
% In this paper, we employ the same space-local filtering in the Navier--Stokes equations to investigate the spatial locality of turbulence.

\begin{comment}
Recently, the combined analysis of both position in space and scale has been conducted using the K\'arm\'an--Howarth--Monin--Hill equation~\citep{Valente2015_the_energy, Yasuda2018_spatio-temporal, Dubrulle2019_beyond}.
Tanogami and Sasa considered an XY spin model with space-local interactions as a toy model of turbulence.
They reported an alternative universality class in the energy spectrum for two- and three-dimensions~\citep{Tanogami2022_XY}.
\end{comment}

% ++++++++++++++++++++++++++++++++++++++++++++++++++
% \clearpage
% \subsection*{Approach and results}
% ++++++++++++++++++++++++++++++++++++++++++++++++++

% % --------------------------------------------------
% \subsubsection*{Space-local Navier--Stokes turbulence}
% % --------------------------------------------------

In the present study, we investigate the physical-space locality of three-dimensional turbulence.
To this end, we employ the decomposition proposed by~\citet{Hamlington2008_local}; we define the space-local velocity field by truncating the integral in the Biot--Savart law at radius \(R\) so that the velocity field is induced only by the nearby vorticity field.
Instead of the post-process analysis of the turbulence dataset as done in these studies, we conduct DNS of the modified Navier--Stokes equations in which the space-local velocity defines the nonlinear term.
This space-local system simulates a flow with restricted nonlinear interactions, and its behavior is qualitatively different from the post-process analysis of the original Navier--Stokes system.

We anticipate here the results that are obtained in the following.
In the inertial range \(k_f \ll k \ll k_{\eta^>}\), where \(k_f\) and \(k_{\eta^>}\) are the forcing wavenumber and the Kolmogorov wavenumber defined by the high-pass filtered energy dissipation rate [see~\eqref{eq:def_diss_highpass} in \S~\ref{subsec:Scaling for scales smaller than the space-local filter size}], respectively, we observe two regimes.
In the scales smaller than the length scale of the space-local domain \(2\pi/R \ll k \ll k_{\eta^>}\), an \(E(k) \propto k^{-5/3}\) scaling of the energy spectrum is observed.
In the larger scales \(k_f \ll k \ll 2\pi/R\), the enstrophy amplification is suppressed, and the system has a conservative enstrophy cascade in the asymptotic limit of \(R \searrow 0\).
This asymptotic behavior reminds us of turbulence without vortex stretching~\citep{Bos2021_three-dimensional, Wu2022_cascades} with \(E(k) \propto k^{-3}\) scaling.
These observations indicate that:
i) the turbulent energy cascade is sustained by space-local nonlinear interactions and
ii) the absence of spatially nonlocal nonlinearity suppresses large-scale enstrophy amplification, resulting in a conservative enstrophy cascade in the asymptotic limit.

% ++++++++++++++++++++++++++++++++++++++++++++++++++
% \clearpage
% \subsection*{Construction of the paper}
% ++++++++++++++++++++++++++++++++++++++++++++++++++

The paper is constructed as follows.
In \S~\ref{sec:Space-local Navier--Stokes equations}, we discuss the physical space-locality of the Navier--Stokes equations and define its variant: the space-local Navier--Stokes equations with restricted nonlinearity in a space-local sense.
In \S~\ref{sec:Space-local Navier--Stokes turbulence}, we show results of DNS of the space-local Navier--Stokes equations.
We investigate two different scaling regimes of the energy spectrum for larger and smaller than the space-local domain size, respectively.
Using the enstrophy budget equation, we develop a theoretical argument explaining the alternative scaling on large scales.
We conclude the study in \S~\ref{sec:Conclusion} with several perspectives for future investigations.

% ==================================================
% \clearpage
\section{
  Space-local Navier--Stokes equations
  \label{sec:Space-local Navier--Stokes equations}
}
% ==================================================

% ++++++++++++++++++++++++++++++++++++++++++++++++++
\subsection{
  Space locality of the vorticity equation and space-local velocity field
  \label{subsec:Space locality of the vorticity equation and space-local velocity field}
}
% ++++++++++++++++++++++++++++++++++++++++++++++++++

We consider the curl of the incompressible three-dimensional Navier--Stokes equations,
\begin{equation}
  \pdv{\vb*{\omega}}{t}
    + \vb*{u} \vdot \grad{\vb*{\omega}}
    = \vb*{\omega} \vdot \grad{\vb*{u}}
    + \nu \laplacian \vb*{\omega}
    + \grad \times \vb*{f},
  \label{eq:vorticity_eq}
\end{equation}
where \(\vb*{u}\) and \(\vb*{\omega} \equiv \grad \times \vb*{u}\) are the velocity and vorticity fields, respectively.
The forcing field is denoted by \(\vb*{f}\), and the kinematic viscosity \(\nu\) is the control parameter of the system.
The velocity can be obtained from the vorticity by the Biot--Savart law,
\begin{equation}
  \vb*{u} (\vb*{x})
    = \frac{1}{4\pi} \int_\Omega \dd[3]{\vb*{x}'} \frac{\vb*{\omega} (\vb*{x}') \times \qty(\vb*{x} - \vb*{x}')}{\abs{\vb*{x} - \vb*{x}'}^3},
  \label{eq:Biot_Savart_real}
\end{equation}
where \(\Omega\) denotes the entire flow domain or the \(\mathbb{R}^3\) space.
This equation illustrates the spatially nonlocal relation between the velocity and vorticity fields.
The temporal evolution of vorticity at a given point is described by the advection and stretching due to the velocity induced by the \emph{whole} vorticity field, along with viscous damping and forcing.

% ++++++++++++++++++++++++++++++++++++++++++++++++++
% \clearpage
% \subsection{
%   Space-local velocity field
%   \label{subsec:Space-local velocity field}
% }
% ++++++++++++++++++++++++++++++++++++++++++++++++++

Here, we define the space-local velocity field,
\begin{equation}
  \vb*{u}^\mathrm{L} (\vb*{x})
    \equiv \frac{1}{4\pi} \int_{r \le R} \dd[3]{\vb*{x}'} \frac{\vb*{\omega} (\vb*{x}') \times \qty(\vb*{x} - \vb*{x}')}{\abs{\vb*{x} - \vb*{x}'}^3},
  \label{eq:uL_real}
\end{equation}
where \(r \equiv \abs{\vb*{x} - \vb*{x}'}\).
Equation~\eqref{eq:uL_real} involves truncation of the spatial integral of the Biot--Savart law~\eqref{eq:Biot_Savart_real} at a sphere of radius \(R\) centered at \(\vb*{x}\).
Its counterpart is the space-nonlocal velocity field
\begin{equation}
  \vb*{u}^\mathrm{NL} (\vb*{x})
    \equiv \frac{1}{4\pi} \int_{r > R} \dd[3]{\vb*{x}'} \frac{\vb*{\omega} (\vb*{x}') \times \qty(\vb*{x} - \vb*{x}')}{\abs{\vb*{x} - \vb*{x}'}^3},
  \label{eq:uNL_real}
\end{equation}
and we have indeed \(\vb*{u} = \vb*{u}^\mathrm{L} + \vb*{u}^\mathrm{NL}\).
Expression~\eqref{eq:uL_real} was first proposed to investigate the alignment between the vorticity vector and the eigenvectors of the strain-rate tensor~\citep{Hamlington2008_local, Hamlington2008_direct, Hamlington2009_physics-based}.
Recently, an alternative expression of~\eqref{eq:uL_real} in Fourier space was proposed~\citep{Buaria2020_self-attenuation}:
\begin{empheq}[left=\empheqlbrace]{align}
  \vb*{u}^\mathrm{L} (\vb*{k})
    &\equiv \qty[1 - B (kR)] \vb*{u} (\vb*{k}),
  \label{eq:uL_Fourier} \\
  B (kR)
    &= \frac{3 \qty[\sin \qty(kR) - \qty(kR) \cos \qty(kR)]}{\qty(kR)^3},
  \label{eq:fkR_sinc}
\end{empheq}
where \(\vb*{u}(\vb*{k})\) denotes the Fourier transform of \(\vb*{u}(\vb*{x})\), with \(\vb*{k}\) the wavevector and \(k = \abs{\vb*{k}}\).

% ++++++++++++++++++++++++++++++++++++++++++++++++++
% \clearpage
\subsection{
  Space-local Navier--Stokes equations
  \label{subsec:Space-local Navier--Stokes equations}
}
% ++++++++++++++++++++++++++++++++++++++++++++++++++

Using the space-local velocity field~(\ref{eq:uL_real}--\ref{eq:fkR_sinc}), we here define the modified Navier--Stokes equations with restricted nonlinearity in the space-local sense.
The space-local vorticity equation becomes
\begin{equation}
  \pdv{\vb*{\omega}}{t}
    + \vb*{u}^\mathrm{L} \vdot \grad{\vb*{\omega}}
    = \vb*{\omega} \vdot \grad{\vb*{u}^\mathrm{L}}
    + \nu \laplacian \vb*{\omega}
    + \grad \times \vb*{f},
  \label{eq:space-local_vorticity_eq}
\end{equation}
where the velocity \(\vb*{u}\) in the nonlinear terms is substituted by the space-local one \(\vb*{u}^\mathrm{L}\).
This equation is space-locally closed, as the evolution of \(\vb*{\omega}\) at point \(\vb*{x}\) is described by \(\vb*{u}^\mathrm{L}\), locally determined in the sphere of radius \(R\) centered at \(\vb*{x}\).

We here examine the basic properties of~\eqref{eq:space-local_vorticity_eq}.
First, we remark that the space-local system remains incompressible; namely, \(\div{\vb*{u}} = 0\).
Here, \(\vb*{u}\) is governed by the space-local Navier--Stokes equations
\begin{equation}
  \pdv{\vb*{u}}{t}
    = - \vb*{\omega} \times \vb*{u}^\mathrm{L}
    - \grad(p + \frac{\vb*{u}^2}{2})
    + \nu \laplacian \vb*{u}
    + \vb*{f},
  \label{eq:space-local_NSeq}
\end{equation}
corresponding to~\eqref{eq:space-local_vorticity_eq}.
The incompressibility of \(\vb*{u}\) immediately follows from that of the space-local velocity field,
\begin{equation}
  \div{\vb*{u}^\mathrm{L}} = 0
  \quad \leftrightarrow \quad
  \mathrm{i} k_j u_j^\mathrm{L} = \mathrm{i} \qty[1 - B (kR)] k_j u_j = 0.
  \label{eq:uL_incompressibility}
\end{equation}
% Note that the pressure term remains in the space-local Navier--Stokes equations~\eqref{eq:space-local_NSeq}, and this term ensures the incompressibility.

Second, the space-local system violates the Galilean invariance of the original Navier--Stokes equations.
It follows from the property of the space-local filter function~\eqref{eq:fkR_sinc}:
\begin{equation}
  \lim_{k \searrow 0} \qty[1 - B (kR)] = 0,
  \label{eq:space-local_function_in_kto0_limit}
\end{equation}
stating that the \(k = 0\) mode flow is purely nonlocal and is eliminated by filtering.
% This property might be an obstacle to analyzing turbulence with a base flow, such as wall turbulence.
% However, since our current forcing setup~\eqref{eq:def_3DTG} does not generate such \(k = 0\) mode mean flow, this defect is not critical to the following analysis.

Third, by taking the inner product of~\eqref{eq:space-local_NSeq} and \(\vb*{u}\) and integrating over space, we obtain the energy equation,
\begin{equation}
  \dv{E}{t}
    = P
    - \epsilon^\mathrm{NL}
    - \epsilon,
  \label{eq:space-local_energy_equation}
\end{equation}
where
\begin{equation}
  E \equiv \frac12 \int \dd{\vb*{x}} \vb*{u}^2,\quad
  P \equiv \int \dd{\vb*{x}} \vb*{f} \vdot \vb*{u},\quad
  \epsilon \equiv \nu \int \dd{\vb*{x}} \qty(\grad{\vb*{u}})^2,
\end{equation}
and
\begin{equation}
  \epsilon^\mathrm{NL} \equiv \int \dd{\vb*{x}} \qty[\vb*{\omega} \times \vb*{u}^\mathrm{L}] \vdot \vb*{u}.
  \label{eq:def_epsNL}
\end{equation}
% Note that we do not employ the superscript \(\mathrm{L}\) here to distinguish the actual terms of the space-local system from the space-local contributions of the original system.
Here, and in the following, \(\int \dd{\vb*{x}}\) denotes an integral over \(\mathbb{R}^3\).
In our simulations, this integral is evaluated as an integral over the periodic numerical domain.
The additional term \(\epsilon^\mathrm{NL}\) is associated with the spatially nonlocal contributions of the nonlinear term.
It disappears in the \(R \nearrow \infty\) limit, since
\begin{equation}
  \lim_{R \nearrow \infty} \epsilon^\mathrm{NL}
    = \int \dd{\vb*{x}} \qty[\vb*{\omega} \times \vb*{u}] \vdot \vb*{u}
    = 0.
  \label{eq:epsNL_to_0}
\end{equation}

% ==================================================
% \clearpage
\section{
  Space-local Navier--Stokes turbulence
  \label{sec:Space-local Navier--Stokes turbulence}
}
% ==================================================

% ++++++++++++++++++++++++++++++++++++++++++++++++++
\subsection{
  DNS of the space-local Navier--Stokes equations
  \label{subsec:DNS of the space-local Navier--Stokes equations}
}
% ++++++++++++++++++++++++++++++++++++++++++++++++++

We conduct DNS of the space-local Navier--Stokes equations~\eqref{eq:space-local_NSeq} in a triply periodic cube of size \(2\pi\).
See Appendix~\ref{app:Numerical evaluation of the space-local nonlinear term} for the numerical implementation of the space-local nonlinear term.
Throughout this manuscript, we report the result of the three-dimensional Taylor--Green forcing
\begin{equation}
  \vb*{f} = \mqty(
      -f_0 \sin x \cos y \cos z \\
      +f_0 \cos x \sin y \cos z \\
      0
    ),
  \label{eq:def_3DTG}
\end{equation}
with the forcing coefficient \(f_0 = 1\).
See Fig.~\ref{fig:3D_TG_forcing_vortex_structures}(a) in Appendix~\ref{app:Temporal evolution of space-local Navier--Stokes turbulence} for the visualization of the large-scale spherical structures generated by~\eqref{eq:def_3DTG}.
We define the characteristic length scale
\begin{equation}
  L_f \equiv 2\pi/\abs{\vb*{k}_f} = 2\pi/\sqrt{3}
  \label{eq:def_Lf_3DTG}
\end{equation}
and the characteristic time scale
\begin{equation}
  T_f = 1/\sqrt{\abs{\vb*{k}_f} f_0} = 1/\sqrt[4]{3}
  \label{eq:def_Tf_3DTG}
\end{equation}
of the forcing, respectively.
Here \(\vb*{k}_f \equiv \qty(\pm 1, \pm 1, \pm 1)^\intercal\) denotes the forced wave vector where the superscript \(\intercal\) denotes the matrix transpose.

\begin{table}
  \begin{tabular}{cccccccc}
    $N$ & $\nu$ & $u'$ & $\lambda$ & $\epsilon$ & $k_\mathrm{max} \eta$ & $\Re_\lambda$ & $T_\textrm{total} / T_f$ \\ \hline
    % $128$ & $4 \times 10^{-3}$ & \textcolor{red}{\(\ast\ast\ast\)} & \textcolor{red}{\(\ast\ast\ast\)} & \textcolor{red}{\(\ast\ast\ast\)} & \textcolor{red}{\(\ast\ast\ast\)} & \textcolor{red}{\(\ast\ast\ast\)} & \textcolor{red}{\(\ast\ast\ast\)} \\
    $512$ & $6 \times 10^{-4}$ & \(0.783\) & \(0.113\) & \(0.433\) & \(1.14\) & \(147\) & \(81.6\)
  \end{tabular}
  \caption{
    DNS setting and statistical quantities of the DNS of the original Navier--Stokes equations.
    The parameters are the resolution of computational domain \(N\) and kinematic viscosity \(\nu\).
    The statistical quantities are evaluated by time average of:
    the fluctuating isotropic RMS velocity \(u'(t) \equiv \sqrt{2K'(t)/3}\), which is defined by the fluctuating energy \(K'(t) \equiv \expval{u'_i u'_i} / 2 \) where \(\expval{\cdot}\) denotes the spatial average and \(u'_i(\vb*{x}, t)\) denotes the temporal fluctuating velocity field;
    % The temporal fluctuating velocity is defined by \(u'_i (\vb*{x}, t) \equiv u_i (\vb*{x}, t) - \expval{u_i}_t(\vb*{x})\);
    the Taylor microscale \(\lambda(t) \equiv u'(t) \sqrt{15 \nu / \epsilon(t)}\) where the energy dissipation rate is evaluated by \(\epsilon(t) = \nu \expval{\omega_i \omega_i}\);
    the Taylor-length Reynolds number \(\Re_\lambda(t) \equiv u'(t) \lambda(t) / \nu\);
    the simulation time in the statistically steady state \(T_\textrm{total}\) as a function of \(T_f\).
  }
  \label{tab:DNS_setting_statistics_original}
\end{table}

First, we conduct a DNS of developed turbulent flow governed by the original Navier--Stokes equations.
See Fig.~\ref{fig:3D_TG_forcing_vortex_structures}(b) in Appendix~\ref{app:Temporal evolution of space-local Navier--Stokes turbulence} for the visualization of the vortical structures.
Table~\ref{tab:DNS_setting_statistics_original} summarizes the DNS setting and the statistical quantities.

\begin{figure}
  \includegraphics[width=\textwidth]{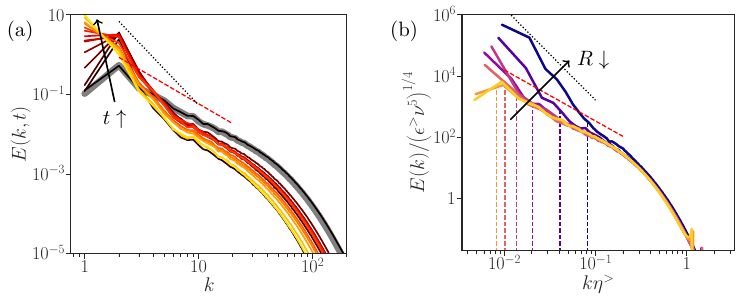}
  \caption[Energy spectrum of space-local turbulence.]{
    (a) Time evolution of energy spectrum \(E(k, t)\) of space-local Navier--Stokes turbulence at \(R = 0.2 L_f\).
    Time evolves from dark to light colors with an interval of \(5T_f\) up to \(70T_f\).
    The thick grey curve represents the time-averaged energy spectrum of the original Navier--Stokes turbulence.
    (b) Normalized instantaneous energy spectrum \(E(k)\) at different values of \(R\).
    Normalization is performed by the high-pass filtered energy dissipation rate \(\epsilon^>\)~\eqref{eq:def_diss_highpass} and the corresponding Kolmogorov length scale \(\eta^>\)~\eqref{eq:def_eta_highpass}.
    Vertical dashed lines denote \(2\pi/R\) normalized by the Kolmogorov length of the original turbulence.
    The darker (lighter) color represents smaller (larger) values of \(R\): \(R/L_f = 0.1,\> 0.2,\> 0.4,\> 0.6,\> 0.8,\> 1\), and the lightest color represents the original turbulence.
    For both panels, the red dashed and black dotted lines denote the \(k^{-5/3}\) and \(k^{-3}\) scalings, respectively.
  }
  \label{fig:energy_spectrum_localNS}
\end{figure}

Then, we launch the DNS of flow governed by the space-local Navier--Stokes equations from a snapshot of this developed turbulent flow.
Figure~\ref{fig:energy_spectrum_localNS}(a) shows the temporal evolution of the energy spectrum \(E(k)\) of turbulence with \(R = 0.2 L_f\).
At first, energy accumulates at \(k = k_f\), and the small scales (\(k \gg k_f\)) become less energetic.
These changes are due to the sudden reduction of the energy cascade.
Another interesting observation is an eventual accumulation of energy in scales larger than the forcing (\(k < k_f\)), suggesting a possible inverse energy cascade.
However, since a further analysis of this property requires a sufficient scale separation between the system size and the forcing scale, we focus on the smaller-scale (\(k > k_f\)) scaling regime, which seems to consist of three ranges:
a range with the energy spectrum with a power law steeper than $k^{-5/3}$,
an \(E(k) \propto k^{-5/3}\) scaling range associated with the Kolmogorov similarity, and
a dissipation range.
See Appendix~\ref{app:Temporal evolution of space-local Navier--Stokes turbulence} for more DNS details, including the time series and visualization.
In the next two subsections, we focus on the first two scaling regimes.

% ++++++++++++++++++++++++++++++++++++++++++++++++++
\subsection{
  \(E(k) \propto k^{-5/3}\) scaling for scales smaller than the space-local filter size
  \label{subsec:Scaling for scales smaller than the space-local filter size}
}
% ++++++++++++++++++++++++++++++++++++++++++++++++++

To investigate the \(R\)-dependence of these regimes, we plot the instantaneous energy spectrum at different values of \(R\) in Fig.~\ref{fig:energy_spectrum_localNS}(b).
The flows are evaluated after the transient when the energy cascade adapts to the truncation of the nonlinear interactions.
See Fig.~\ref{fig:timeseries_localNS} of Appendix~\ref{app:Temporal evolution of space-local Navier--Stokes turbulence} for an alternative definition of the statistically steady state.
We use the same snapshots in the remainder of this article.
In Fig.~\ref{fig:energy_spectrum_localNS}(b), the spectra are normalized by the high-pass filtered energy dissipation rate
\begin{equation}
  \epsilon^> \equiv 2 \nu \int_{2\pi/R}^\infty k^2 E(k) \dd{k},
  \label{eq:def_diss_highpass}
\end{equation}
for which we remove the contributions from the \emph{direct} energy dissipation in wavenumber range \(k < 2\pi/R\) without energy cascade.
Accordingly, the modified Kolmogorov length
\begin{equation}
  \eta^> \equiv \qty(\nu^3 / \epsilon^>)^{1/4}
  \label{eq:def_eta_highpass}
\end{equation}
is used to normalize the wavenumber.

% --------------------------------------------------
% \subsubsection*{Robust K41 universality}
% --------------------------------------------------

In the small scales \(2\pi/R \ll k\), the normalized energy spectra collapse onto the Kolmogorov spectrum.
We also observe that \(E(k) \propto k^{-5/3}\) law in \(2\pi/R \ll k \ll k_{\eta^>} (= 2\pi/\eta^>)\) is robust.
Here, \(2\pi/R\) is the wavenumber corresponding to the space-local domain of radius \(R\).
This result indicates that the system with only spatially local nonlinear interactions (parametrized by \(R\)) can sustain energy cascade in scales smaller than \(R\).

We stress the nontriviality of this result; the space-local structures in physical space within a sphere of radius \(R\) are not equivalent to the small-scale structures in Fourier space for \(k \ge 2\pi/R\).
Similarly, the space-local domain of radius \(R\) contains (partial) information of all the Fourier modes, not only \(k \ge 2\pi/R\).
Thus, there is no one-to-one correspondence between the space-local and the small-scale structures.
Overall, Fig.~\ref{fig:energy_spectrum_localNS}(b) confirms that the observations are consistent with the original space-local assumption in~\citet{Kolmogorov1941_the_local}.

Here, we note that we investigate the range of \(0.1 \le R/L_f \le 1\).
Since \(L_f > \pi\) with the current forcing configuration~\eqref{eq:def_3DTG}, the largest space-local domain exceeds the computational domain of \((2\pi)^3\) and we cannot correctly calculate the distance \(r\).
However, its effect is rather small in Fig.~\ref{fig:energy_spectrum_localNS}(b).
In this sense, we state that the numerical results show the convergence of the space-local Navier--Stokes turbulence to the original Navier--Stokes turbulence in the \(R \nearrow \infty\) limit.

% ++++++++++++++++++++++++++++++++++++++++++++++++++
% \clearpage
\subsection{
  Enstrophy-conserving scaling in scales larger than the space-local filter size
  \label{subsec:Enstrophy-conserving scaling in scales larger than the space-local filter size}
}
% ++++++++++++++++++++++++++++++++++++++++++++++++++

To understand the steeper scaling of the spectrum in \(k_f \ll k \ll 2\pi/R\), we investigate the enstrophy balance in space-local Navier--Stokes turbulence.
We consider the large-scale enstrophy budget equation
\begin{equation}
  \pdv{t} \int_0^k p^2 E(p) \dd{p}
    + \varPi_\omega (k)
    = V_\omega^<(k)
    - \epsilon_\omega^< (k)
    + \int_0^k F_\omega(p) \dd{p}
  \label{eq:cumulative_enstrophy_eq_Fourier}
\end{equation}
of the Navier--Stokes equations.
In three-dimensional isotropic turbulence, there is a balance between the cumulative enstrophy amplification \(V_\omega^<(k)\) and the enstrophy flux \(\varPi_\omega(k)\) in the inertial range~\citep{Davidson2008_on_the_generation, Sadhukhan2019_enstrophy}.
These terms correspond to the vortex stretching and advection terms of the vorticity equation~\eqref{eq:space-local_vorticity_eq}.
For the definition, derivation, and scaling of~\eqref{eq:cumulative_enstrophy_eq_Fourier}, see Appendix~\ref{app:Enstrophy balance and its scaling}.

\begin{figure}
  \includegraphics[width=0.6\textwidth]{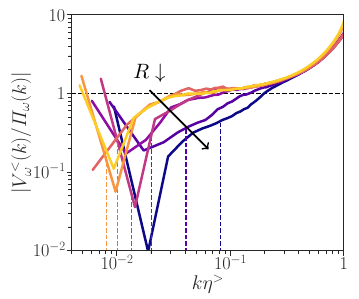}
  \caption[enstrophy amplification/flux balance in space-local NS turbulence.]{
    Ratio of cumulative enstrophy amplification \(V_\omega^<(k)\) and enstrophy flux \(\varPi_\omega(k)\).
    The horizontal dashed line denotes \(V_\omega^<(k) = \varPi_\omega(k)\).
    The same colormap and the vertical dashed lines are employed as in Fig.~\ref{fig:energy_spectrum_localNS}(b).
  }
  \label{fig:enstrophy_production_flux_balance_localNS}
\end{figure}

Figure~\ref{fig:enstrophy_production_flux_balance_localNS} shows the ratio \(V_\omega^< (k) / \varPi_\omega (k)\) between the cumulative enstrophy amplification rate \(V_\omega^< (k)\) and the enstrophy flux \(\varPi_\omega (k)\).
These two terms are balanced in the inertial range of the unmodified Navier--Stokes turbulence (corresponding to the \(R \nearrow \infty\) limit), and \(V_\omega^< (k) / \varPi_\omega (k)\) should thus be unity.
See also Fig.~\ref{fig:enstrophy_budget} of Appendix~\ref{app:Enstrophy balance and its scaling}.
In space-local Navier--Stokes turbulence, the ratio becomes considerably smaller than unity for \(k_f \ll k \ll 2\pi/R\) as \(R \searrow 0\), which is, as we will argue now, associated with the suppression of enstrophy amplification.

We consider the global enstrophy amplification rate,
\begin{equation}
  V_\omega
    = \expval{\omega_i \pdv{u_i^\mathrm{L} (\vb*{x})}{x_j} \omega_j} \\
    = \expval{\omega_i \omega_j \pdv{x_j} \mathcal{F}^{-1} \qty[u_i^\mathrm{L} (\vb*{k})]},
    \label{eq:def_enstrophy_production_rate}
\end{equation}
of space-local Navier--Stokes turbulence.
Here, \(\mathcal{F}^{-1} \qty[\cdot]\) and \(\expval{\cdot}\) denote the inverse Fourier transform and the spatial average, respectively.
In the limit of \(kR \ll 1\), a Taylor expansion of the space-local filter function~\eqref{eq:fkR_sinc} yields
\begin{equation}
  1 - B(kR)
    \approx \frac{(kR)^2}{10} + \order{(kR)^3}.
  \label{eq:space_local_filter_kR<<1_limit}
\end{equation}
Thus, in this limit, the total enstrophy amplification rate~\eqref{eq:def_enstrophy_production_rate} scales as
\begin{empheq}{align}
  V_\omega
    % &\approx \expval{\omega_i \omega_j \pdv{x_j} \mathcal{F}^{-1} \qty[\frac{(kR)^2}{10} u_i (\vb*{k})]} \\
    &\approx \frac{R^2}{10} \expval{\omega_i \omega_j \pdv{x_j} \mathcal{F}^{-1} \qty[k^2 u_i (\vb*{k})]}
    \qfor kR \ll 1.
  \label{eq:enstrophy_production_rate_kRll1}
\end{empheq}
This relation states that the enstrophy amplification at a given scale \(k\) is weakened by decreasing the radius of the space-local domain \(R\), and is consistent with Fig.~\ref{fig:enstrophy_production_flux_balance_localNS}.

From these numerical and theoretical observations, we conjecture that space-local Navier--Stokes turbulence is asymptotically equivalent to turbulence without vortex stretching~\citep{Bos2021_three-dimensional, Wu2022_cascades} in the \(kR \ll 1\) limit.
This system has intermediate properties between two- and three-dimensional Navier--Stokes turbulence, as enstrophy and helicity are conserved in the inviscid limit.
Note that the nonlinearity of the two-dimensional Navier--Stokes equations conserves energy and enstrophy, while the three-dimensional system conserves energy and helicity.

\begin{figure}
  \includegraphics[width=\textwidth]{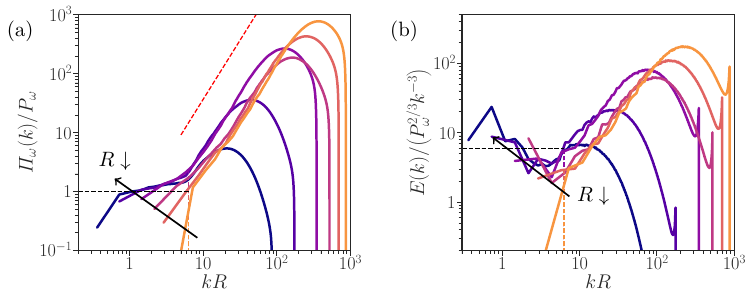}
  \caption[Enstrophy flux and compensated energy spectrum of space-local turbulence.]{
    (a) Enstrophy flux \(\varPi_\omega(k)\) normalized by the enstrophy input rate \(P_\omega\).
    The red dashed line denotes \(k^2\) scaling.
    (b) Compensated energy spectrum \(E(k)\) based on the scaling~\eqref{eq:Ek_space-local_k-3}.
    For both panels, the wavenumber is normalized by \(R\) so that different \(2\pi/R\) collapses onto \(kR = 2\pi\).
    Note that the spectrum of the original turbulence is not shown, since it corresponds to the \(R \nearrow \infty\) limit.
    The horizontal dashed line denotes a plateau.
    The same colormap is employed as in Fig.~\ref{fig:energy_spectrum_localNS}(b).
  }
  \label{fig:enstrophy_flux_compensated_energy_spectrum_localNS}
\end{figure}

We numerically verify this speculation in Fig.~\ref{fig:enstrophy_flux_compensated_energy_spectrum_localNS}.
We first show the enstrophy flux \(\varPi_\omega (k)\) normalized by the enstrophy injection rate
\begin{equation}
  P_\omega
    \equiv \expval{\qty(\nabla \times \vb*{f}) \vdot \vb*{\omega}}
    % = \expval{\qty(\epsilon_{ijk} \partial_j f_k) \omega_i},
  \label{eq:enstrophy_injection_rate}
\end{equation}
in Fig.~\ref{fig:enstrophy_flux_compensated_energy_spectrum_localNS}(a).
The wavenumber is normalized by \(R\) so that the flux is horizontally shifted.
Under the space-local constraint, we observe shallower scaling in the large scales \(k \ll 2\pi/R\), indicating a constant enstrophy flux \(\varPi_\omega (k) \propto k^0\) in the asymptotic limit of \(R \searrow 0\).
Since there is no enstrophy amplification in this limit, the magnitude of \(\varPi_\omega (k)\) in this regime is of the order of the total enstrophy injection \(P_\omega\) by the forcing~\eqref{eq:enstrophy_injection_rate}.

In an enstrophy-conserving system, the energy spectrum exhibits an asymptotic scaling of \(E(k) \propto k^{-3}\),
% \begin{equation}
%   E(k) \propto k^{-3},
%   \label{eq:Ek_wo_vortex_stretching}
% \end{equation}
for example, see~\citet[Fig.~1]{Bos2021_three-dimensional}.
Figure~\ref{fig:enstrophy_flux_compensated_energy_spectrum_localNS}(b) shows the compensated energy spectrum according to
\begin{equation}
  E(k) \sim P_\omega^{2/3} k^{-3},
  \label{eq:Ek_space-local_k-3}
\end{equation}
associated with the conservative enstrophy cascade picture.
Although we do not observe a clear plateau even for the smallest value of the \(R = 0.1 L_f\) snapshot, the large-scale behavior in \(R \searrow 0\) limit is not contradict the asymptotic enstrophy conservation discussed in Fig.~\ref{fig:enstrophy_flux_compensated_energy_spectrum_localNS}(a).
Plausibly, we may observe clearer \(k^{-3}\) scaling with large enough separation between \(k_f\) and \(2\pi/R\), which would require much larger computational capacity.
Furthermore, logarithmic corrections can also affect this scaling~\citep{Kraichnan1971_inertial, Wu2022_cascades}.

% ==================================================
% \clearpage
\section{
  Conclusion
  \label{sec:Conclusion}
}
% ==================================================

% Research objective
The scale locality of nonlinear interactions in three-dimensional turbulence has received considerable attention in the turbulence community, unlike locality in physical space.
The present study aims to understand how spatially local and nonlocal nonlinear interactions contribute to the small-scale universality of turbulence.
To this end, we considered the space-local velocity field~\eqref{eq:uL_real} induced by the \emph{space-local} vorticity field.
Here, \emph{space-local} is defined by the contributions contained in a spherical region of radius \(R\) around the considered point in space.
We use this velocity to define a variant of the vorticity equation~\eqref{eq:space-local_vorticity_eq} in which the nonlinear term is determined in the space-local sense.

\begin{figure}
  \includegraphics[width=0.8\textwidth]{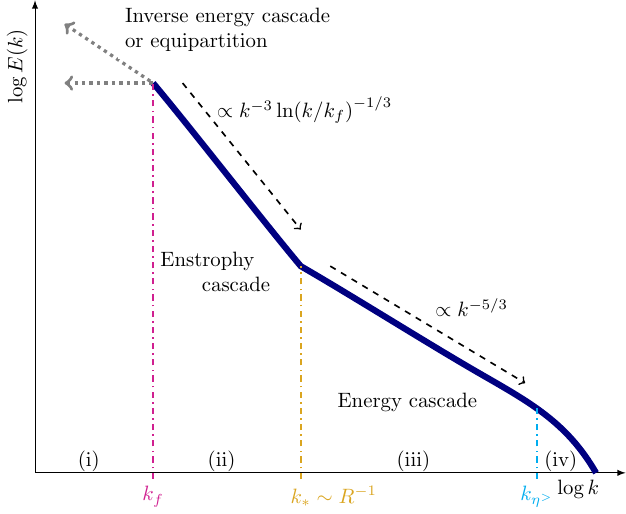}
  \caption[A schematic of space-local Navier--Stokes turbulence.]{
    A schematic of the energy spectrum associated with space-local Navier--Stokes turbulence.
    Dashed arrows denote \(k^{-3}\) (plus logarithmic correction) and \(k^{-5/3}\) scaling, respectively.
    Vertical dash-dotted lines denote: the forcing wavenumber \(k_f\), the intersecting wavenumber \(k_\ast\) of the two scalings, and the Kolmogorov wavenumber \(k_{\eta^>}\), respectively.
    Annotations (i)-(iv) correspond to different scaling regions divided by the characteristic wavenumbers.
  }
  \label{fig:schematic_space-local_NS_turbulence}
\end{figure}

% Results
We conducted DNS of this space-local flow in a \(2\pi\)-periodic box driven by the steady forcing~\eqref{eq:def_3DTG}.
The wavenumber range of the energy spectrum \(E(k)\) of space-local Navier--Stokes turbulence can be decomposed into the following regions:
(i) a possible inversely energy cascading range in \(k \ll k_f\),
(ii) \(E(k) \propto k^{-3}\) scaling range associated with a conservative enstrophy cascade in the asymptotic limit of \(R \searrow 0\) in \(k_f \ll k \ll 2\pi/R\),
(iii) \(E(k) \propto k^{-5/3}\) scaling range in \(2\pi \ll k \ll k_{\eta^>}\), and
(iv) a dissipation range.
Here, \(k_f, 2\pi/R,\) and \(k_{\eta^>}\) denote the characteristic wavenumbers of the forcing, physical-space locality, and dissipation, respectively.
These regimes are schematically summarized in Fig.~\ref{fig:schematic_space-local_NS_turbulence}.

% Significance
When we focus on the inertial range \(k_f \ll k \ll k_{\eta^>}\) [regions (ii) and (iii)], the space-local Navier--Stokes equations consolidate the robustness of Kolmogorov similarity with energy cascade in its small-scale part [regions (iii) and (iv)].
This finding suggests that the nonlinear interactions of three-dimensional turbulence are local in physical space as well as in scale space.
The physical-space locality of the nonlinear interactions is consistent with \citet{Kolmogorov1941_the_local}'s hypothesis, where spatially local domain was considered.
The large-scale part [region (ii)] behaves asymptotically as turbulence without vortex stretching, which is explained by the suppressed enstrophy amplification and corresponds to a constant enstrophy flux.
We note that this spectral shape with two (asymptotic) scaling ranges is similar to the Nastrom--Gage spectrum of atmospheric turbulence~\citep{Nastrom1984_kinetic}, first reported in the late 1960s~\citep{Wiin1967_on_the_annual}.
In that case, the enstrophy-conserving range with the \(k^{-3}\) scaling corresponds to close to two-dimensional turbulence while the \(k^{-5/3}\) scaling is recovered in the small scales.

% Perspectives
Extensive investigations with higher resolution and wider scaling range between \(k_f\) and \(2\pi/R\) are needed to confirm the \(E(k) \propto k^{-3}\) scaling and its intersection with the \(E(k) \propto k^{-5/3}\) scaling.
Furthermore, the nature of a possible inverse cascade of the injected energy with small finite values of \(R\), as in Fig.~\ref{fig:energy_spectrum_localNS}(b), is not investigated in the current study.
Indeed, three-dimensional turbulence can exhibit inverse cascades if the nonlinear term is modified~\citep{Biferale2012_inverse, Frisch2012_turbulence, Wu2022_cascades}.
An alternative configuration with a much larger scale separation between the system size and the forcing would make it possible to investigate the behavior of the system in the \(k \ll k_f\) regime.
Two-dimensional space-local turbulence may exhibit qualitatively different properties compared to the three-dimensional case since the two-dimensional Navier--Stokes equations are governed by long-range (spatially nonlocal) interactions.
Recent investigations show that space locality is important in the dynamics of the large-scale condensation in two-dimensional turbulence~\citep{Svirsky2023_two-dimensional}.

% \clearpage
\appendix
\section{
  Numerical evaluation of the space-local nonlinear term
  \label{app:Numerical evaluation of the space-local nonlinear term}
}
% ==================================================

This appendix presents the numerical implementation of the space-local nonlinear term.
The space-local vorticity equation reads, without forcing and damping,
\begin{equation}
  \pdv{\vb*{\omega}(\vb*{x})}{t}
    = - \grad \times \qty(\vb*{\omega} \times \vb*{u}^\mathrm{L}) (\vb*{x}),
  \label{eq:inviscid_vorticity}
\end{equation}
and its Fourier transform is
\begin{equation}
  \pdv{\vb*{\omega}(\vb*{k})}{t}
    = - \mathrm{i} \vb*{k} \times \qty(\vb*{\omega} \times \vb*{u}^\mathrm{L}) (\vb*{k}).
  \label{eq:inviscid_vorticity_Fourier}
\end{equation}
We can retrieve the Euler equations by uncurling~\eqref{eq:inviscid_vorticity_Fourier} as,
\begin{empheq}{align}
  \pdv{\vb*{u}}{t}
    &= - \frac{\mathrm{i} \vb*{k}}{k^2} \times \qty[\mathrm{i} \vb*{k} \times \qty(\vb*{\omega} \times \vb*{u}^\mathrm{L})], \\
    &= - P_{ij} \qty(\vb*{\omega} \times \vb*{u}^\mathrm{L})_j,
\end{empheq}
where \(P_{ij} = \delta_{ij} - k_i k_j / k^2\).
Using this formulation, the space-local nonlinear term in the Fourier domain can be computed using standard pseudo-spectral procedures.

% ==================================================
% \clearpage
\section{
  Temporal evolution of space-local Navier--Stokes turbulence
  \label{app:Temporal evolution of space-local Navier--Stokes turbulence}
}
% ==================================================

\begin{figure}
  \includegraphics[width=0.8\textwidth]{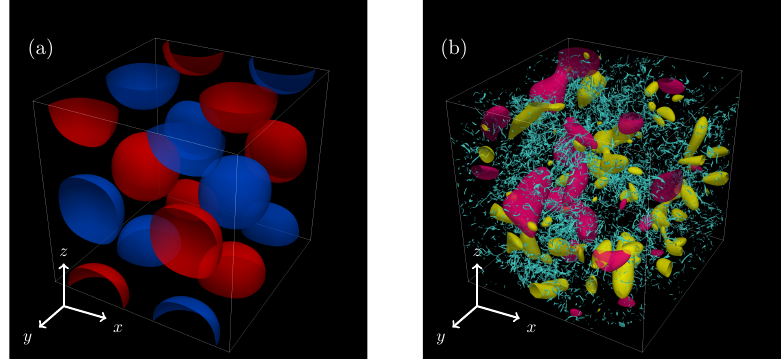}
  \caption[Visualization of the forcing and the vortical structures of the original turbulence.]{
    (a) Positive (red) and negative (blue) isosurfaces of \(x\)-component of the forcing field, \(f_x = \pm 0.5\).
    See~\eqref{eq:def_3DTG} for the definition.
    (b) An instantaneous snapshot of vortical structures.
    Isosurfaces of vorticity magnitude in low-pass filtered \(\abs{\vb*{\omega}^<} = 4\) for \(k \le 3\) (red), band-pass filtered \(\abs{\vb*{\omega}^\lessgtr} = 6\) for \(3 < k \le 6\) (yellow), and original \(\abs{\vb*{\omega}} = 100\) (blue) are shown.
    Low-pass and band-pass filtering are applied in the Fourier-space velocity field.
  }
  \label{fig:3D_TG_forcing_vortex_structures}
\end{figure}

This appendix investigates the dynamics and structures in the space-local Navier--Stokes turbulence.
Before this, we first look at those in the original Navier--Stokes turbulence.
Figure~\ref{fig:3D_TG_forcing_vortex_structures}(a) visualizes positive and negative isosurfaces of the \(x\)-component of the forcing field~\eqref{eq:def_3DTG}.
It consists of large-scale spherical objects, which correspond to the vortical structures in the steady flow at low Reynolds numbers as well as the largest vortices in turbulence at high Reynolds numbers.
Figure~\ref{fig:3D_TG_forcing_vortex_structures}(b) shows the vortical structures in a snapshot of developed turbulence generated by the original Navier--Stokes equations.
The observations are consistent with higher Reynolds number turbulence visualizations presented, for example, in~\citet{Goto2017_hierarchy}.

\begin{figure}
  \includegraphics[width=\textwidth]{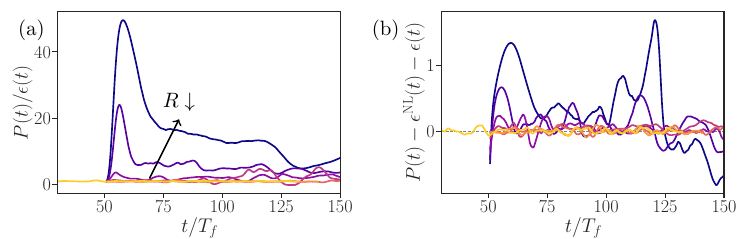}
  \caption[Time series of energy input rate and nonlocal dissipation of space-local turbulence.]{
    Time series of (a) energy input rate \(P(t)\) normalized by the energy dissipation rate \(\epsilon(t)\) and (b) RHS of the energy equation~\eqref{eq:space-local_energy_equation} of space-local Navier--Stokes turbulence at different values of \(R\).
    For both panels, time is normalized by \(T_f\): the characteristic timescale of the forcing~\eqref{eq:def_Tf_3DTG}.
    The horizontal grey dashed line in panel (b) denotes \(y=0\).
    See the caption of Fig.~\ref{fig:energy_spectrum_localNS}(b) for the parameter list for \(R\) corresponding to each curve with a different color.
  }
  \label{fig:timeseries_localNS}
\end{figure}

Figure~\ref{fig:timeseries_localNS}(a) shows the time series of \(P(t) / \epsilon(t)\), the ratio between input and dissipation rates of energy.
By switching the governing equation from original to space-local Navier--Stokes equations, the energy input rate surpasses the energy dissipation rate due to the weakened nonlinearity and energy cascade.
This transient regime appears as a peak in the time series.
We observe a higher peak for flows with smaller values of \(R\) because the nonlinearity is more suppressed for small \(R\).

After the transient stage, the flow reaches a state where \(P(t) / \epsilon(t)\) seems to fluctuate around a constant value which depends on \(R\).
It therefore differs from a statistically steady state of the original Navier--Stokes turbulence, as there is no statistical balance between the injection and the dissipation of energy: \(\expval{P(t)}_t \ne \expval{\epsilon(t)}_t\).
Here, \(\expval{\cdot}_t\) denotes the time average.
In Fig.~\ref{fig:timeseries_localNS}(b), we plot \(P(t) - \epsilon^\mathrm{NL}(t) - \epsilon(t)\), i.e.~the RHS of the energy equation~\eqref{eq:space-local_energy_equation}, to assess the temporal evolution of the space-local Navier--Stokes turbulence.
For the original Navier--Stokes turbulence, we observe \(\expval{P(t) - \epsilon^\mathrm{NL}(t) - \epsilon(t)}_t = 0\) with \(\epsilon^\mathrm{NL} = 0\)~\eqref{eq:epsNL_to_0}.
When the value of \(R\) is finite, we still observe fluctuations around zero after the initial transient regime depicted in Fig.~\ref{fig:timeseries_localNS}(a).
This observation indicates that the space-local Navier--Stokes turbulence eventually establishes the statistically steady state, as the three terms on the RHS of the energy equation~\eqref{eq:space-local_energy_equation} attain a statistical balance and thus \(\expval{\dv*{E(t)}{t}}_t = 0\).

We note that for the smallest value of \(R/L_f = 0.1\), we observe a large fluctuation amplitude indicating a significant instantaneous imbalance between the three terms.
However, since there are both positive and negative values, we speculate that there is a statistical balance in the long enough time series.
Unfortunately, due to the energy accumulation at large scales, long enough computation to evaluate the statistical convergence is beyond our computational capacity.

\begin{comment}
Figure~\ref{fig:timeseries_localNS}(b) plots the time series of \(\epsilon^\mathrm{NL}(t)\), illustrating that \(\epsilon^\mathrm{NL} \to 0\) for \(R \nearrow \infty\) as discussed in~\eqref{eq:epsNL_to_0} of \S~\ref{subsec:Space locality of the vorticity equation and space-local velocity field}.
Note that \(\epsilon^\mathrm{NL}(t)\) does not always act as an alternative energy dissipation as it can be negative.
\end{comment}

\begin{figure}
  \includegraphics[width=0.8\textwidth]{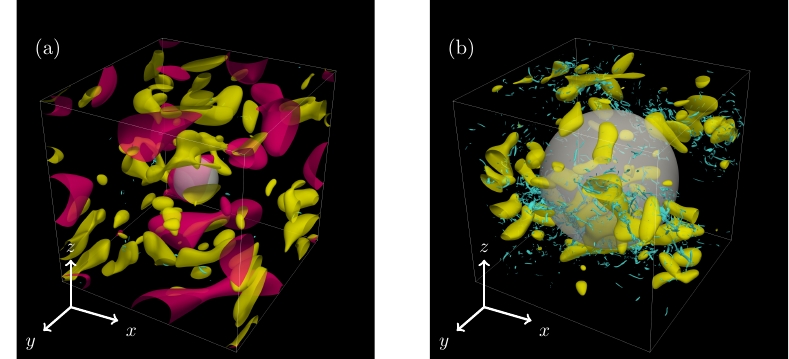}
  \caption[Visualization of the vortical structures of space-local turbulence.]{
    Snapshots of vortical structures in space-local Navier--Stokes turbulence at (a) \(R = 0.2 L_f\) and (b) \(R = 0.6 L_f\), respectively.
    Isosurfaces of the magnitude of low-pass filtered vorticity \(\abs{\vb*{\omega}^<} = 8\) (red), band-pass filtered one \(\abs{\vb*{\omega}^\lessgtr} = 3\) (yellow), and unfiltered one \(\abs{\vb*{\omega}} = 50\) (blue) are shown.
    See the caption of Fig.~\ref{fig:3D_TG_forcing_vortex_structures}(b) for the filtering wavenumber ranges.
    Note that we employ smaller thresholds compared to the visualization of the original turbulent flow in Fig.~\ref{fig:3D_TG_forcing_vortex_structures}(b), but they are common in the panels (a) and (b) of this figure.
    The grey spherical domain illustrates the size of the locality parameter \(R\) at the center of the computational domain.
  }
  \label{fig:Visualization_localNS}
\end{figure}

Figure~\ref{fig:Visualization_localNS} shows the vortical structures in space-local Navier--Stokes turbulence at \(R = 0.2 L_f\) and \(R = 0.6 L_f\).
By comparing them with original Navier--Stokes turbulence in Fig.~\ref{fig:3D_TG_forcing_vortex_structures}(b), we can see that the space-local turbulence has much less fine-scale structures (in blue), even though smaller thresholds are employed for the visualization.
It supports our finding that the nonlinear interactions and energy cascade are weakened due to the space-local restrictions in the nonlinear term.

Next, we compare the two panels of Fig.~\ref{fig:Visualization_localNS} with the same isosurface thresholds.
For \(R = 0.2 L_f\) in panel~(a), there are distinctive large-scale structures (in red and yellow), while the small-scale structures (in blue) are barely visible.
This indicates that more energy remains at these large-scale structures because less energy cascades towards the scales smaller than \(R\).
For \(R = 0.6 L_f\) in panel~(b), we do not observe the strong large-scale structures (in red), whereas the small-scale structures are more active.
This is also consistent with the picture that less energy is retained at large scales as the energy cascade becomes more efficient for larger \(R\) because the space-local restrictions become less significant.
More quantitative arguments are developed in the main text in terms of the energy spectrum shown in Fig.~\ref{fig:energy_spectrum_localNS}.

% ==================================================
% \clearpage
\section{
  Enstrophy balance and its scaling
  \label{app:Enstrophy balance and its scaling}
}
% ==================================================

This appendix investigates the enstrophy balance scaling of the original Navier--Stokes turbulence in Fourier space.
We begin with the enstrophy balance equation
\begin{equation}
  \pdv{t} k^2 E(k)
    = T_\omega(k)
    + S_\omega(k)
    - 2 \nu k^4 E(k)
    + F_\omega(k),
  \label{eq:enstrophy_eq_Fourier}
\end{equation}
where \(k^2 E(k)\) denotes the enstrophy spectrum.
There are two nonlinear terms, namely, the enstrophy amplification
\begin{equation}
  S_\omega (k)
    = \int \qty(\omega_j  \partial_j u_i) (\vb*{k}) \omega_i^\ast (\vb*{k}) \dd{\Omega_k}
  \label{eq:def_enstrophy_production}
\end{equation}
and the enstrophy transfer
\begin{equation}
  T_\omega (k)
    = \int -\qty(u_j \partial_j \omega_i) (\vb*{k}) \omega_i^\ast (\vb*{k}) \dd{\Omega_k},
  \label{eq:def_enstrophy_transfer}
\end{equation}
associated with the stretching and advection term in the vorticity equation~\eqref{eq:space-local_vorticity_eq}, respectively.
Here, \(\cdot^\ast\) and \(\int \dd{\Omega_k}\) denote the complex conjugate and the integral over spherical shells of radius \(k\), respectively.
The third term on the RHS denotes the enstrophy dissipation, and the fourth term is the enstrophy injection
\begin{equation}
  F_\omega (k)
    = \int \qty(\epsilon_{ijk} \partial_j f_k) (\vb*{k}) \omega_i^\ast (\vb*{k}) \dd{\Omega_k}.
  \label{eq:def_enstrophy_injection}
\end{equation}

\begin{figure}
  \includegraphics[width=0.6\textwidth]{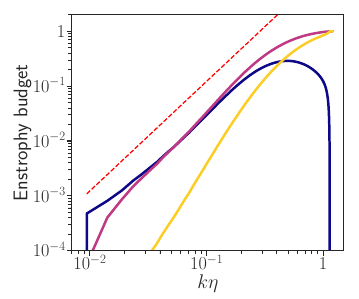}
  \caption[enstrophy amplification/flux/dissipation.]{
    Three terms from the enstrophy budget equation~\eqref{eq:app:cumulative_enstrophy_eq_Fourier}: enstrophy flux \(\varPi_\omega(k)\), cumulative enstrophy amplification \(V_\omega^<(k)\), and cumulative enstrophy dissipation rate \(\epsilon_\omega(k)\).
    Each spectrum and the wavenumber are normalized by the enstrophy dissipation rate \(\epsilon_\omega^< = 2 \nu \int_0^\infty k^4 E(k) \dd{k}\) and the Kolmogorov length scale \(\eta\), respectively.
    Vertical solid and dashed lines denote \(k_f\) and \(k_\eta\), respectively.
  }
  \label{fig:enstrophy_budget}
\end{figure}

The large-scale integral of~\eqref{eq:enstrophy_eq_Fourier} defines the enstrophy budget equation
\begin{equation}
  \pdv{t} \int_0^k p^2 E(p) \dd{p}
    + \varPi_\omega (k)
    = V_\omega^<(k)
    - \epsilon_\omega^< (k)
    + \int_0^k F_\omega(p) \dd{p},
  \label{eq:app:cumulative_enstrophy_eq_Fourier}
\end{equation}
which is equivalent to~\eqref{eq:cumulative_enstrophy_eq_Fourier}.
Figure~\ref{fig:enstrophy_budget} shows the three terms of~\eqref{eq:app:cumulative_enstrophy_eq_Fourier}: the cumulative enstrophy amplification
\begin{equation}
  V_\omega^< (k)
    = \int_0^k S_\omega (p) \dd{p},
  \label{eq:def_cumulative_enstrophy_production}
\end{equation}
the enstrophy flux
\begin{equation}
  \varPi_\omega (k)
    = - \int_0^k T_\omega (p) \dd{p},
  \label{eq:def_enstrophy_flux}
\end{equation}
and the cumulative enstrophy dissipation
\begin{equation}
  \epsilon_\omega^< (k)
    = 2 \nu \int_0^k p^4 E(p) \dd{p}.
  \label{eq:def_cumulative_enstrophy_dissipaton}
\end{equation}
In the inertial range \(k_f \ll k \ll k_\eta (= 2\pi/\eta)\), we observe a balanced scaling of
\begin{equation}
  \varPi_\omega(k)
    = V_\omega^<(k)
    \propto k^2,
  \label{eq:enstrophy_production_flux_scaling}
\end{equation}
associated with the plateau in the energy flux \(\varPi(k) \propto k^0\)~\citep{Davidson2008_on_the_generation, Sadhukhan2019_enstrophy}.
Given that \(T(k) \propto - \pdv*{\varPi(k)}{k}\), where \(T(k)\) is the energy transfer, \(V_\omega^< (k)\) and \(\varPi_\omega (k)\) can be expressed as
\begin{empheq}{align}
  -k^2 \pdv{\varPi}{k}
    = -\pdv{k} \qty[k^2 \varPi(k)]
    + 2k \varPi(k)
    = T_\omega(k)
    + V_\omega^<(k).
\end{empheq}
The inertial range of the enstrophy budget equation~\eqref{eq:app:cumulative_enstrophy_eq_Fourier}, shown in Fig.~\ref{fig:enstrophy_budget}, can be understood as the scale-by-scale balance between the cumulative enstrophy amplification and enstrophy flux~\eqref{eq:enstrophy_production_flux_scaling}.
Although no conservative enstrophy cascade exists, enstrophy is transferred from larger to smaller scales.
More precisely, the enstrophy transferred to scale \(k\) from a larger scale is further transferred towards a smaller scale, along with the enstrophy generated at that scale.

In Fig.~\ref{fig:enstrophy_production_flux_balance_localNS}, we show the balance of \(V_\omega^< (k) / \varPi_\omega (k)\) for both the original and space-local Navier--Stokes turbulence.
For the latter, the enstrophy amplification and transfer are defined by
\begin{empheq}[left=\empheqlbrace]{align}
  S_\omega (k)
    &= \int \qty(\omega_j  \partial_j u^\mathrm{L}_i) (\vb*{k}) \omega_i^\ast (\vb*{k}) \dd{\Omega_k},
  \label{eq:def_enstrophy_production_localNS} \\
  T_\omega (k)
    &= \int -\qty(u^\mathrm{L}_j \partial_j \omega_i) (\vb*{k}) \omega_i^\ast (\vb*{k}) \dd{\Omega_k},
  \label{eq:def_enstrophy_transfer_localNS}
\end{empheq}
where \(\vb*{u}\) in~\eqref{eq:def_enstrophy_production} and~\eqref{eq:def_enstrophy_transfer} is substituted by \(\vb*{u}^\mathrm{L}\).

% \clearpage
\begin{acknowledgments}
  R.~A. thanks Dr.~Remi Zamanski for his insightful comments, which helped us to develop the discussion in \S~\ref{subsec:Space-local Navier--Stokes equations}.
  R.~A. also thanks Dr.~Genta Kawahara, Dr.~Kosuke Osawa and Dr.~Tomohiro Tanogami for their comments and discussions.
  R.~A and W.~B appreciate the discussion with Dr.~Alain Pumir.
  R.~A. and S.~G. appreciate the discussion with Dr.~Tsuyoshi Yoneda.

  All DNS calculations were conducted in the facilities of the PMCS2I (\'Ecole Centrale de Lyon).
  R.~A. is supported by the Takenaka Scholarship Foundation.

  For the purpose of Open Access, a CC-BY public copyright license has been applied by the authors to the present document and will be applied to all subsequent versions up to the Author Accepted Manuscript arising from this submission.
\end{acknowledgments}

\bibliography{reference}

\end{document}